\documentclass{rspublic}
\usepackage{graphicx}

\begin{document}

\title[Shock break-out]{Shock break-out: how a GRB revealed the beginnings of a supernova}

\author[A. J. Blustin]{{Alexander J. Blustin}\thanks{On behalf of the Swift Team}}

\affiliation{UCL Mullard Space Science Laboratory, Holmbury St. Mary, Dorking, Surrey RH5 6NT, UK}

\label{firstpage}

\maketitle

\begin{abstract}{gamma rays: bursts; supernovae: individual (SN 2006aj); stars: Wolf-Rayet; shock waves}
In February 2006, Swift caught a GRB in the act of turning into a supernova, and made the first ever direct observations of the break-out and early expansion of a supernova shock wave. GRB 060218 began with an exceptionally long burst of non-thermal gamma-rays, lasting over 2000~s, as a jet erupted through the surface of the star. While this was in progress, an optically-thick thermal component from the shock wave of the supernova explosion grew to prominence, and we were able to track the mildly relativistic expansion of this shell as the blackbody peak moved from the X-rays into the UV and optical bands. The initial radius of the shock implied that it was a blue supergiant which had exploded, but the lack of Hydrogen emission lines in the supernova spectrum indicated a more compact star. The most likely scenario is that the shock ploughed into the massive stellar wind of a Wolf-Rayet progenitor, with the shock breaking out and becoming visible to us once it reached the radius where the wind became optically-thin. I present the Swift observations of this landmark event, and discuss the new questions and answers it leaves us with.
\end{abstract}

\section{Introduction}

The Gamma-Ray Burst (GRB) detected by the Swift (Gehrels \textit{et al.} 2004) Burst Alert Telescope (BAT; Barthelmy \textit{et al.} 2005) during a slew on 18 February 2006 looked unusual right from the beginning. The initial burst of $\gamma$-rays was both exceptionally long (T$_{\rm 90}$ = 2100 $\pm$ 100~s) and peaked at a very low energy (4.9$^{+0.4}_{-0.3}$~keV), the latter of which classifies the event as an X-Ray Flash (XRF). At the redshift of the event, z=0.03345 $\pm$ 0.0006 (Mirabal et al. 2006) making it the second closest GRB yet detected, its isotropic equivalent energy was E$_{\rm iso}$ = 6 $\times$ 10$^{49}$ erg which is 100 to 1000 times less than most GRBs. By way of comparison, GRB 980425, the first GRB to be identified with a supernova, had a much shorter T$_{\rm 90}$ of 23.3 $\pm$ 1.4~s (Galama \textit{et al.} 1998), a higher E$_{peak}$ of 55 $\pm$ 21~keV, and a lower E$_{\rm iso}$ of (1.0 $\pm$ 0.2) $\times$ 10$^{48}$~erg (Amati 2006).

\begin{figure}
     \includegraphics[width=8cm,angle=-90]{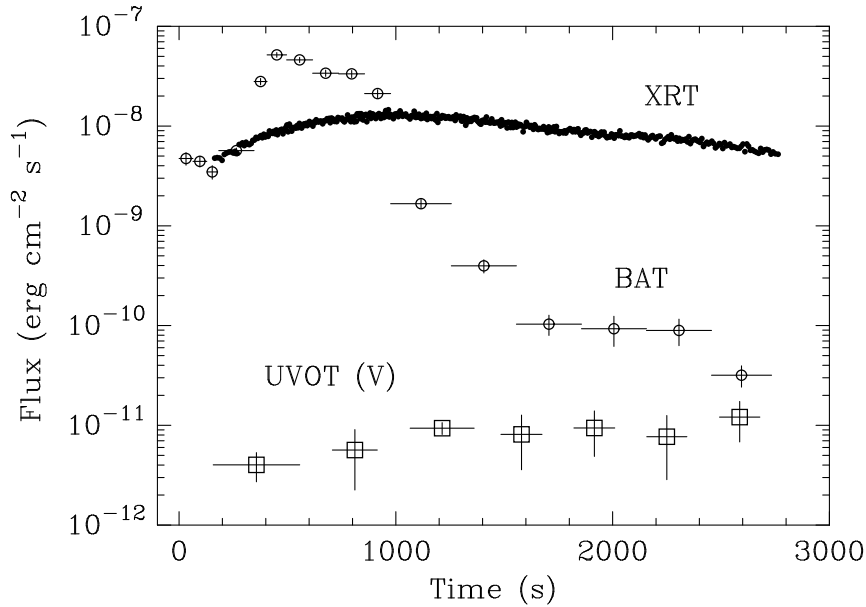}
   \caption{$\gamma$-ray (BAT), X-ray (XRT) and V-band (UVOT) lightcurves from the first orbit ($\sim$ 3000~s) of Swift observations of GRB060218 (after Campana \textit{et al.} 2006). The BAT flux is calculated by fitting XRT and BAT data together, so the BAT flux is dominated by the extrapolation of the XRT fit into the BAT energy range.}
   \label{1storbit}
   \end{figure}

\begin{figure}
     \includegraphics[width=10cm,angle=-90]{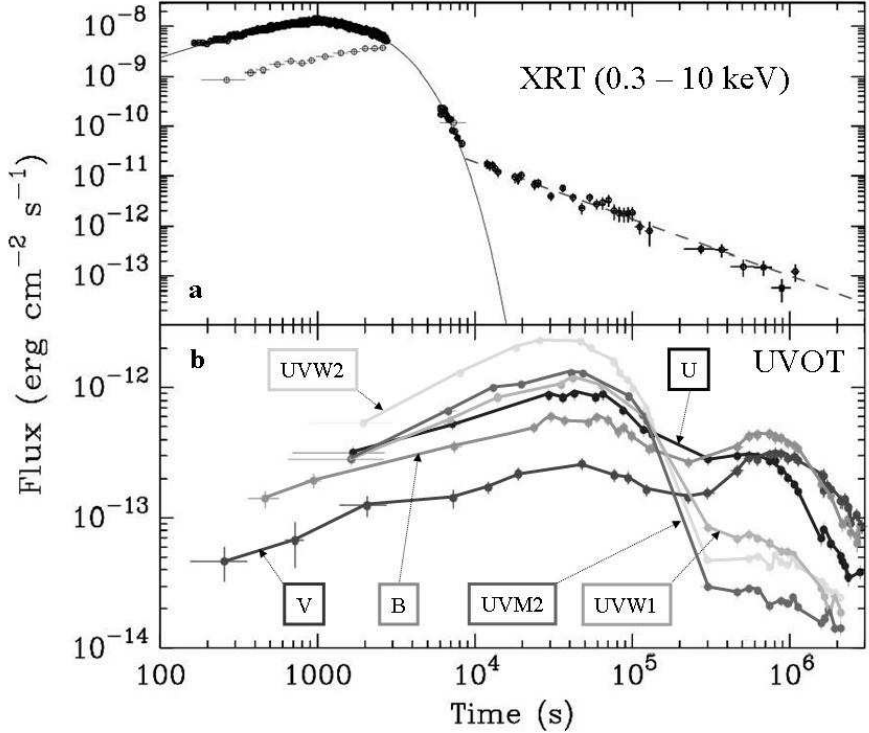}
   \caption{a) X-ray lightcurve of GRB060218, 0.3 to 10 keV, (black circles) with the contribution of the thermal component (grey circles); b) UVOT lightcurves in the six colour filter bands: V (centred at 5440\AA), B (4390\AA), U (3450\AA), UVW1 (2510\AA), UVM2 (2170\AA) and UVW2 (1880\AA) (after Campana \textit{et al.} 2006).}
   \label{total_lcs}
   \end{figure}

During the first 3000~s of the Swift observation of GRB 060218 (figure \ref{1storbit}) the 0.3 to 10~keV X-ray lightcurve from the X-Ray Telescope (XRT; Burrows \textit{et al.} 2005) shows a smooth peak and gradual decay, whilst after 3000~s the optical V-band flux observed with the UV/Optical Telescope (UVOT; Roming \textit{et al.} 2005) is still rising. 10~ks after the beginning of the burst (figure \ref{total_lcs}a), the exponentially decaying X-ray lightcurve gives way to a power-law decay of the type more typically seen in GRB afterglows. The optical/UV flux registered by the UVOT (figure \ref{total_lcs}b) rises to a UV dominated peak at $\sim$ 40~ks, and later on to a peak dominated by optical light at 600~ks. All of this unusual phenomenology offers clues as to the nature of the progenitor and the physical evolution of the event, which I will discuss in the following sections. These observations and their interpretation are presented in greater detail by Campana \textit{et al.} (2006).

\section{The progenitor}

\begin{figure}
     \includegraphics[width=12cm]{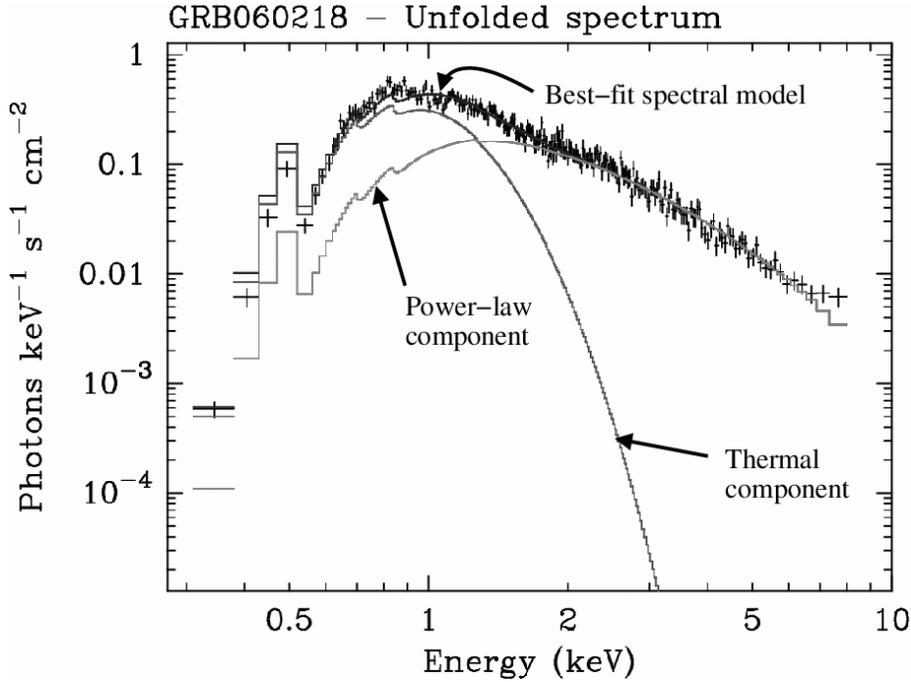}
   \caption{X-ray spectrum of GRB060218, 0.3 to 10 keV, compiled from data taken up to $\sim$ 3 ks after the beginning of the event, with the best fitting spectral model superimposed on the data. The thermal and power-law components are also plotted.}
   \label{xrtspec}
   \end{figure}

The early behaviour of the X-ray emission gives us information about the progenitor. The X-ray spectrum compiled from data taken during the first 3~ks shows (figure \ref{xrtspec}) that there is a thermal component to the emission as well as the power-law typically observed in GRB spectra. The flux in this thermal component evolves with time; its flux is plotted alongside the total X-ray flux in figure \ref{total_lcs}a. The thermal flux dominates the X-ray emission between $\sim$ 2 and 8~ks after the beginning of the event, but dies away exponentially after that point. Fitting a blackbody model to this thermal emission using XSPEC, we obtained a temperature of kT $\approx$ 0.17 keV. The characteristic radius $R_{shell}$ of the emitting region can be estimated as:

\begin{equation} R_{shell} \approx \left(\frac{E_{iso}}{a T^4}\right)^\frac{1}{3} \end{equation}

Where $E_{iso}$ is the equivalent isotropic power of the GRB, $a$ is the radiation density constant and $T$ is the thermal temperature. The resulting radius, $\sim$ 5$\times$10$^{12}$~cm, would imply a blue supergiant progenitor, but the lack of Hydrogen lines in the optical spectra of the supernova (e.g. Mazzali \textit{et al.} 2006) implies a more compact star. The larger apparent radius can, however, be explained by the presence of a massive stellar wind. Assuming that the early expansion of the emitting region was linear, due to light-travel-time effects, we can estimate an initial progenitor radius of (4 $\pm$ 1)$\times$10$^{11}$~cm. This size, and the presence of the optically thick stellar wind, implies that the progenitor was a Wolf-Rayet star.

\section{The shockwave}

The shockwave of the explosion becomes visible when it reaches the radius when the wind becomes optically thin. The shock is probably non-spherical, since the characteristic radius $R_{shell}$, which is calculated on the assumption of spherical symmetry, is larger than the radius of the blackbody as fitted to the XRT data. This may be due to the presence of the presumed GRB jet, or perhaps to the stellar wind being non-spherical. Since we actually observe the process of the shock breaking out of the stellar wind, we can constrain the delay between the GRB and supernova start to be less than 4~ks.

The shock compresses the wind to a thin shell. The optical depth must be about 1, so we can obtain its opacity, estimate the mass of the shell and therefore the wind mass-loss rate $\dot M$:

\begin{equation} \dot M \sim \left(\frac{M_{shell} v_{wind}}{R_{shell}}\right) \end{equation}

where we take $v_{wind}$ = 10$^{8}$~cm~s$^{-1}$, a typical value for Wolf-Rayet stars. We obtain ${\dot M}$ $\sim$ 3$\times$10$^{-4}$~$M_{\odot}$~yr$^{-1}$, and using the radius $R_{shell}$, the wind density $\rho_{wind}$ $\sim$ 10$^{12}$~g~cm$^{-3}$. The thermal energy density behind a radiation-dominated shock is

\begin{equation} a T_{BB}^{4} \sim 3 \rho_{wind} v_{shock}^{2} \end{equation}

where $T_{BB}$ is the blackbody temperature and $v_{shock}$ is the shock speed. Using $\rho_{wind}$ that we have calculated, $v_{shock}$ $\approx$ c.

\section{The stellar remnant}

\begin{figure}
     \includegraphics[width=12cm]{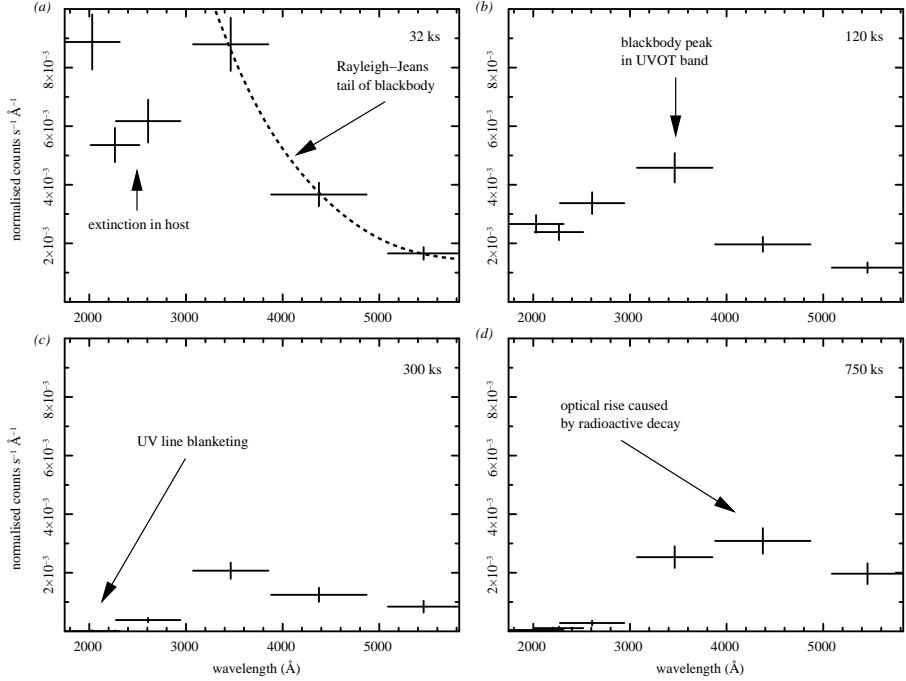}
   \caption{The evolution of the UVOT broad-band spectrum from 32 ks to 750 ks after the beginning of the event. The data points represent the count rate per {\AA}ngstrom in each of the UVOT filters (note that the instrumental response has not been removed in these plots): left to right: UVW2 (centred at 1880\AA), UVM2 (2170\AA), UVW1 (2510\AA), U (3450\AA), B (4390\AA) and V (5440\AA). }
   \label{uvotspec}
   \end{figure}

At the UV emission peak observed with the UVOT, the blackbody emission in the UV and optical is far higher than that expected from the shock. Instead, the emission is probably from the expanding stellar remnant. The expansion of the remnant is traced by the evolution of the broad-band UVOT spectrum, as shown in figure \ref{uvotspec}. At the time of the earliest spectrum in the figure, 32~ks (figure \ref{uvotspec}a), the peak of the blackbody is still far higher than the UVOT band, and so the UVOT spectrum consists of the Rayleigh-Jeans tail of the blackbody which is cut into by intrinsic extinction in the GRB host. This allows us to fit the depth of the extinction, which is then fixed in subsequent fits of the blackbody parameters. Using the \emph{zdust} model in XSPEC, we obtain an intrinsic E(B-V) = 0.20 $\pm$ 0.03, assuming an SMC reddening law (Pei, 1992).

One hundred and twenty kiloseconds after the beginning of the event (figure \ref{uvotspec}b), the blackbody peak is centred in the UVOT band and so its temperature and radius can be fitted. We obtain kT $\sim$ 3.7$^{+1.9}_{-0.9}$ eV and $R_{BB}$ = 3.29$^{+0.94}_{-0.93}\times10^{14}$~cm. This radius implies an expansion speed of 27000 $\pm$ 8000~km~s$^{-1}$, consistent with the speeds obtained from optical spectroscopy (e.g. Sollerman \textit{et al.} 2006, Pian \textit{et al.} 2006, Mazzali \textit{et al.} 2006). By 300~ks (figure \ref{uvotspec}c), the remnant has cooled to the extent that elements can combine, and so the UV flux is drastically decreased from that point on by line blanketing. Finally, at 750~ks (figure \ref{uvotspec}d), the optical band has re-brightened due to radioactive decay in the supernova remnant.

\section{Conclusions}

Our conclusions are, then, that the progenitor of GRB060218/SN2006aj was a Wolf-Rayet star, and that the one-of-a-kind lightcurves obtained with Swift signify the breakout of the shockwave from the dense stellar wind of the progenitor, and the expansion and cooling of the stellar remnant. 

This unusual event was followed up by many other telescopes and observers, leading to a range of novel possibilities and conclusions. Firstly, the probable $\sim$20~M$_\odot$ mass of the progenitor was low enough for a magnetar remnant --rather than a black hole-- to have formed, leading to the intriguing suggestion that XRFs are the magnetic activity of magnetars forming from the implosion of low mass progenitors, whilst Gamma-Ray Bursts signal the collapse of high mass stars and the formation of black holes (Mazzali \textit{et al.} 2006, Soderberg \textit{et al.} 2006). The presence or otherwise of relativistic ejecta or a collimated jet may also be a fundamental distinction between supernovae accompanied by GRBs, and those which are not (Soderberg \textit{et al.} 2006, Pian \textit{et al.} 2006). Whatever the compact remnant is, the long duration of the afterglow is a sign of continuing activity of a rapidly spinning central engine (Soderberg \textit{et al.} 2006, Fan \textit{et al.} 2006). The fact that this low-luminosity event was even detected implies that there is a much larger number of such events than previously thought; perhaps, even, that they are the most common high-energy transient events in the Universe (Cobb \textit{et al.} 2006, Pian \textit{et al.} 2006).

\end{document}